# Optically Induced Picosecond Lattice Compression in the Dielectric Component of a Strongly Coupled Ferroelectric/Dielectric Superlattice

*Deepankar Sri Gyan, Hyeon Jun Lee, Youngjun Ahn, Jerome Carnis, Tae Yeon Kim, Sanjith Unithrattil, Jun Young Lee, Sae Hwan Chun, Sunam Kim, Intae Eom, Minseok Kim, Sang-Youn Park, Kyung Sook Kim, Ho Nyung Lee, Ji Young Jo, and Paul G. Evans**

D. Sri Gyan, Dr. H. J. Lee, Dr. Y. Ahn, Prof. P. G. Evans

Department of Materials Science and Engineering, University of Wisconsin-Madison, Madison, Wisconsin 53706, USA

E-mail: pgevans@wisc.edu

Dr. J. Carnis

Aix Marseille Université, Université de Toulon, CNRS, IM2NP, Marseille, France

ESRF - The European Synchrotron, 71 Avenue des Martyrs, 38000 Grenoble, France

Dr. T. Y. Kim, Dr. S. Unithrattil, J. Y. Lee, Prof. J. Y. Jo

School of Materials Science and Engineering, Gwangju Institute of Science and Technology, Gwangju 61005, South Korea

Dr. S. H. Chun, Dr. S. Kim, Dr. I. Eom, Dr. M. Kim, Dr. S. Y. Park, Dr. K. S. Kim

Pohang Accelerator Laboratory, Pohang, Gyeongbuk 37673, South Korea

Dr. H. N. Lee

Materials Science and Technology Division, Oak Ridge National Laboratory, Oak Ridge, Tennessee, 37831, USA

Above-bandgap femtosecond optical excitation of a ferroelectric/dielectric $BaTiO_3/CaTiO_3$ superlattice leads to structural responses that are a consequence of the screening of the strong electrostatic coupling between the component layers. Time-resolved x-ray free-electron laser diffraction shows that the structural response to optical excitation includes a net lattice expansion of the superlattice consistent with depolarization-field screening driven by the photoexcited charge carriers. The depolarization-field-screening-driven expansion is separate from a photoacoustic pulse launched from the bottom electrode on which the superlattice was epitaxially grown. The distribution of diffracted intensity of superlattice x-ray reflections indicates that the depolarization-field-screening-induced strain includes a photoinduced expansion in the ferroelectric $BaTiO_3$ and a contraction in $CaTiO_3$. The magnitude of expansion in $BaTiO_3$ layers is larger than the contraction in $CaTiO_3$. The difference in the magnitude of depolarization-field-screening-driven strain in the $BaTiO_3$ and $CaTiO_3$ components can arise from the contribution of the oxygen octahedral rotation patterns at the $BaTiO_3/CaTiO_3$ interfaces to the polarization of $CaTiO_3$. The depolarization-field-screening-driven polarization reduction in the $CaTiO_3$ layers points to a new direction for the manipulation of polarization in the component layers of a strongly coupled ferroelectric/dielectric superlattice.

## Introduction

Epitaxial superlattice (SL) heterostructures consisting of alternating repeating layers of ferroelectric (FE) and dielectric (DE) complex oxides have polarization configurations that result from a combination of atomic-scale and mesoscopic effects.[1-2] Key effects include the electrostatic coupling between layers, the atomic-scale structure of interfaces, strain arising from the epitaxial mismatch, and the depolarization field arising from the polarization discontinuity at interfaces between FE and DE layers.[3-5] At the nanometer length scale, an internal electric field polarizing the DE layers arises in response to the depolarization field in the FE layer.[6] Structural features such as the octahedral rotation pattern of the component layers can also vary at, and across, interfaces and can affect the equilibrium configuration of the electrical polarization.[2, 7] Non-equilibrium conditions arising after the absorption of an above-bandgap optical pulse further expand the range of phenomena observed in FE/DE SLs. For example, the screening of the depolarization field in SL heterostructures by excited charge carriers leads to transformations between domain configurations and to novel polarization states.[8-13] Charge carriers excited due to the above-bandgap optical excitation of the SL can result in depolarization-field screening, which can reduce the magnitude of the internal electric field polarizing the DE layers. This depolarization-field-screening driven by optical excitation presents a promising way to modulate the polarization states of the FE/DE SL at an ultrafast timescale. Here, we present a structural study probing the response of a FE/DE SL to a femtosecond optical pulse and find that the optically induced screening results in a photoinduced lattice compression of the DE layer.

The equilibrium polarization configuration of FE/DE SLs is determined in large part by the electrostatic interaction between the component layers.[6] The defining issue is that uncompensated charges due to the divergence of the polarization at the FE/DE interface can result in a significant increase in the free energy of the SL.[1] Several possible configurations must be considered as possible routes to reduce relevant free energies. The SL can, for example, adopt a continuous polarization in which there is a decrease in the polarization of the FE layers in comparison with their bulk form and the development of the polarization in the DE layers comparable to that of the FE layers.[14-15] Theoretical studies based on the Landau-Ginzburg-Devonshire (LGD) theory or using density functional theory both predict the emergence of a spontaneous polarization in the DE layers of FE/DE SLs.[6, 16] The octahedral rotation configuration at the FE/DE interfaces can also contribute to the development of electrical polarization in the DE layers.**[17]**

FE/DE SLs can be broadly categorized as strongly or weakly coupled based on the electrostatic coupling between the FE layers. SLs with strong electrostatic coupling favor a highly polarized DE layer, rather than the formation of domains or other complex configurations.[6, 15] In contrast, weakly coupled SLs have lower polarization in the DE layers and can exhibit nanodomain and vortex patterns reducing the electrostatic energy that would result from a discontinuity of ferroelectric polarization arising from the creation of an interface between a fully polarized FE layer and a weakly polarized DE layer.[1, 18-20] A dielectric-slab model based on the LGD theory provides a general guideline for the conditions that determine whether a particular SL heterostructure exhibits strong or weak interlayer coupling, with parameters including the relative thickness of the FE and DE layers.[21]

The high concentration of charge carriers resulting from above-bandgap optical excitation changes the local electrostatic conditions by screening the depolarization field.[22-23] The depolarization-field screening by optically excited charge carriers can simultaneously increase the polarization of FE layers and reduce the polarization in the DE layers. More specifically, above-bandgap optical excitation can be hypothesized to lead to a high concentration of carriers leading to depolarization-field screening and a subsequent lattice distortion coupled to the polarization of the component layers. This hypothesis is illustrated for a strongly coupled $BaTiO_3/CaTiO_3$ (BTO/CTO) SL in **Figure 1a**. Changes in the lattice spacing of the BTO and CTO layers respectively, are expected to result from optical excitation. The magnitudes of the change in polarizations of the two layers, $\Delta P_{BTO}$ and $\Delta P_{CTO}$, are not expected to be equal because the polarization of CTO layers does not depend solely on the depolarization field.

In this work, the photoinduced lattice distortion in a strongly coupled BTO/CTO SL was probed using ultrafast x-ray free-electron-laser (XFEL) diffraction. The experimental arrangement is shown in **Figure 1b**. The SL heterostructure produces several x-ray reflections that include the Bragg reflections arising from the average lattice parameter of the SL and satellite reflections with a wavevector spacing along the surface-normal *z*-direction set by the SL repeating layer thickness. Optically induced structural distortion changes the intensities of the x-ray reflections of the SL, from which the structural changes within the component layers can be determined. The intensities of the SL reflections depend on the relative thicknesses of the BTO and CTO layers and the structure factors of the individual layers and hence can be used to measure the component-specific response.[24-25] The structural responses of the BTO and CTO layers can be determined precisely using a kinematic x-ray diffraction simulation to interpret the observed intensities of the SL x-ray reflections.

The BTO/CTO SL considered here consists of a repeating unit of 2-unit cells (u. c.) of BTO and 4 u. c. of CTO on an $SrRuO_3$ (SRO) bottom electrode on an $SrTiO_3$ (STO) substrate, as illustrated in Figure 1a,b. The SL thin film has an overall thickness of approximately 200 nm. The analysis below considers a model structure consisting of exactly 80 periods corresponding to a thickness $d$=191 nm. SL satellite x-ray reflections are labeled with an integer index $l$, with the 002 SL Bragg reflection at $l$=0, as shown in **Figure 1c**. Satellite reflections at higher and lower values of the out-of-plane scattering vector $Q_z$ with respect to the Bragg reflection at $l$=0 have positive and negative values of $l$, respectively. The dielectric-slab model predicts that the BTO/CTO SL will exhibit a strong coupling for CTO volume fractions less than a critical value of 33%.[21] Despite having a CTO volume fraction of 67%, larger than the critical value, the 2:4 BTO/CTO SL exhibits strong coupling with nearly equal polarizations of the BTO and CTO layers.[25-26] A precise prediction of the polarization of SLs can be obtained by including the effect of structural reconstruction at interfaces.[2, 17, 27-29] The combined effect of electrical, structural, and mechanical boundary conditions imposed by the SL geometry results in a polarization in the CTO layers nearly equal to BTO layers in BTO/CTO SL.[17, 30]

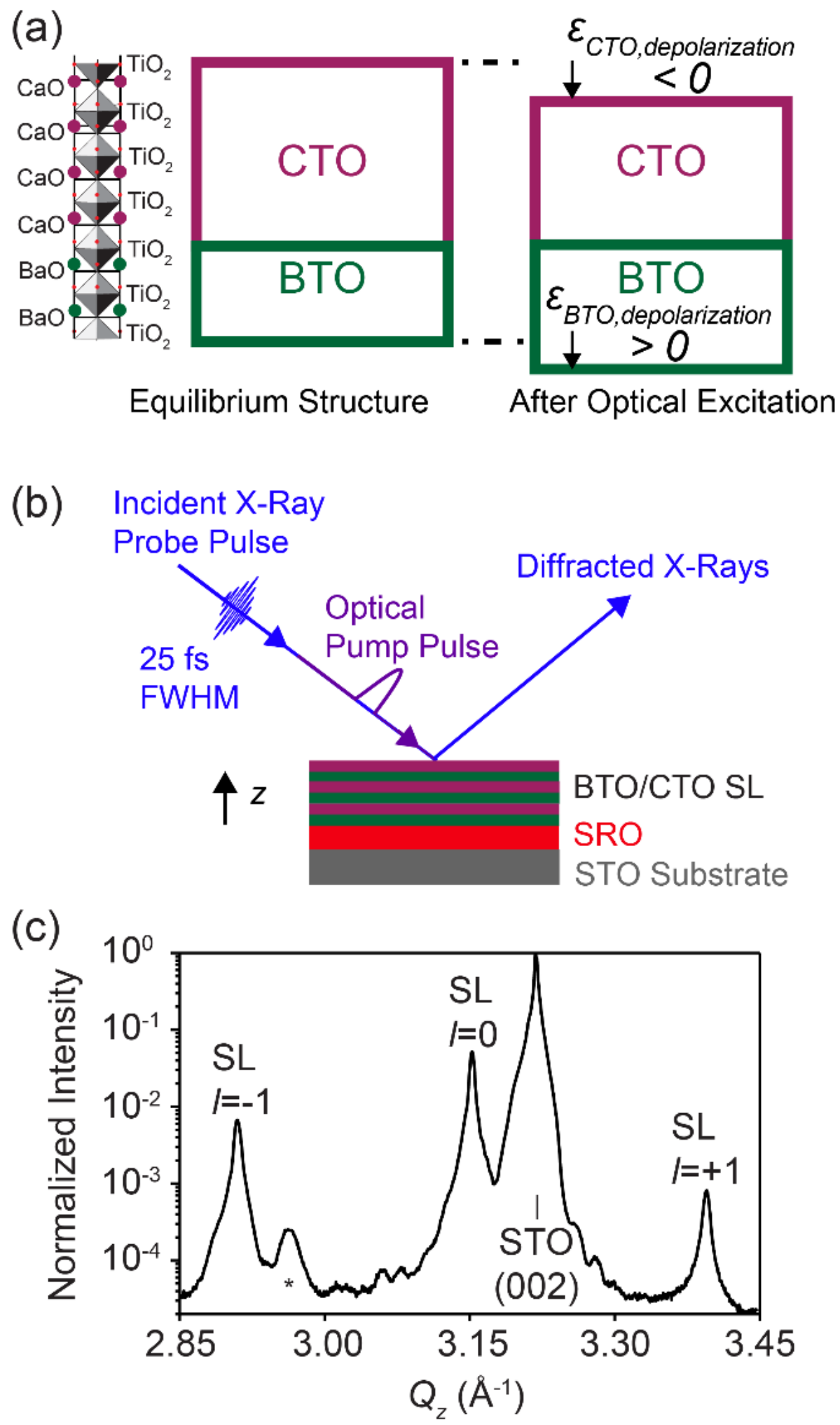

**Figure 1. a)** Atomic arrangement of BTO/CTO SL and schematic of the photoinduced change in the structure arising from the change in the polarization of the component layers. Dashed lines indicate the equilibrium structure before the optical excitation. **b)** Layer structure and experimental arrangement for ultrafast x-ray diffraction study of the optically excited BTO/CTO SL. **c)** Steady-state diffraction pattern acquired with a laboratory x-ray source including the SL *l*=0, +1, and -1 reflections near the 002 Bragg reflection and the STO substrate. The SRO bottom electrode contributes intensity near the STO 002 reflection. An additional reflection from a small fractional

component of the thin film is indicated by * and is not considered in the analysis.

## Results and Discussion

The distortion of the component layers of the SL was determined using the intensities and reciprocal-space locations of the 002 Bragg reflection of the SL, at $l$=0, and the SL satellite reflections with $l$=+1 and $l$=-1. The diffracted x-ray intensities depend on $Q_z$ and on the time interval $t$ between the optical excitation pulse and the x-ray probe pulse. The experimentally measured intensities are shown in **Figure 2a-c**. After optical excitation at $t$=0, each reflection exhibits an initial shift of the diffracted intensity to higher $Q_z$ because of the initial compression of the SL by an acoustic pulse launched from the SRO bottom electrode. At times after 64 ps, there is a shift towards lower $Q_z$, corresponding to a lattice expansion after the propagation of the acoustic pulse into the substrate. A detailed analysis below shows that the responses of the two component layers, however, are different.

The diffraction data in Figure 2 exhibits a series of temporal oscillations of the intensity. The temporal period of the intensity oscillations depends on $\Delta Q_z$, the difference between the value of $Q_z$ and the wavevector of Bragg or SL reflection at each time. The variation of the oscillation period as a function of $\Delta Q_z$ is a signature of the propagation of an acoustic pulse.[13, 31] Analysis of the time dependence of the intensity at the $l$=+1 SL reflection near the 002 reflection and a Fourier analysis of the intensity near this reflection both give a longitudinal sound velocity $v_{SL}$ = 5.9 (±0.2) km $s^{-1}$, as described in the Supporting Information. The acoustic pulse evident in Figure 2 arises because optical absorption heats the conducting SRO layer and establishes a stress profile at $t$=0. The stress in the SRO electrode launches acoustic pulses into the SL and the substrate.[32] The profile and propagation of the strain pulse are discussed in detail in the Supporting Information. The predicted amplitude of the strain pulse due to optical absorption in the SRO layer under the experimental conditions is 0.7%.

Optical absorption directly in the SL also leads to an increase in its temperature. However, as described in the Supporting Information, the amplitude of the acoustic strain pulse arising from the heating of the SL is very small, on the order of 0.001%, in comparison with the 0.7% peak strain of the acoustic strain from the SRO layer. An acoustic pulse is also generated by the stress profile due to depolarization-field screening in the layers, for example through the development of a depolarization-field-screening-driven acoustic pulse in $BiFeO_3$.[33] The pulse due to the depolarization-field screening, on the order of 0.01%, is also far smaller than the amplitude of the acoustic strain pulse from the SRO layer. The distortion in the interval in which the acoustic pulse propagation occurs (0-64 ps) is thus dominated by the pulse generated by the heating of the SRO.

The acoustic pulse from the SRO bottom electrode propagates through the total thickness of the SL in time $\tau = d/v_{SL}$ = 32 ps. The acoustic perturbation is observed up to time $t = 2\tau$ in Figure 2 because the acoustic pulse from the SRO bottom electrode is reflected at the SL/air interface at $t = \tau$ and then propagates toward the SL/SRO interface. The high acoustic impedance mismatch at the SL/air interface causes the acoustic pulse to be reflected from the free surface with a 180° phase change.[34] After $2\tau$ = 64 ps, the strain pulse reaches the SL/SRO interface and propagates through the SRO into the substrate. The low acoustic impedance mismatch at the SL/SRO and SRO/STO interfaces causes only 20% of the acoustic amplitude to be reflected towards the surface, as described in more detail in the Supporting Information.

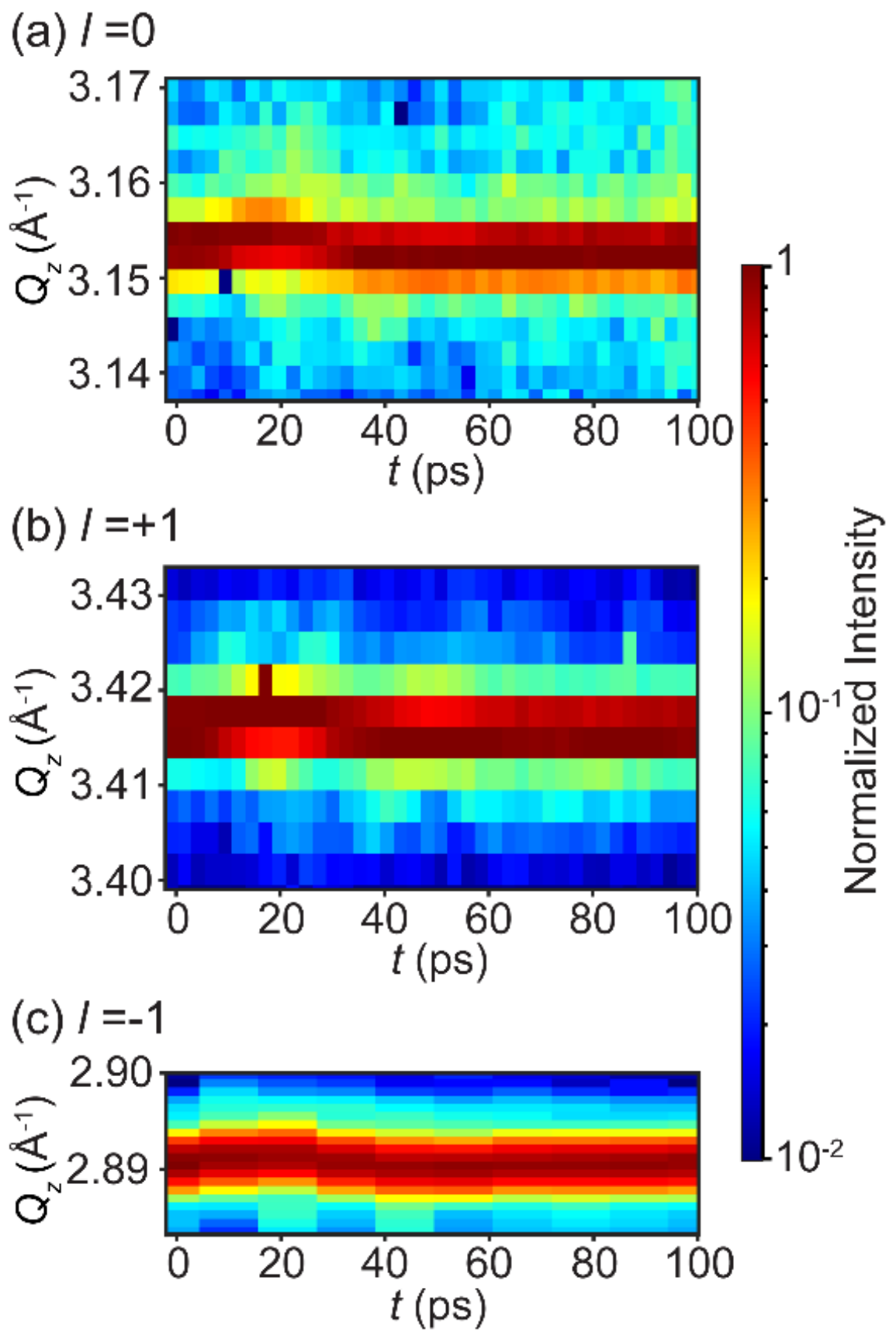

**Figure 2.** Scattered x-ray intensity as a function of $Q_z$ and time $t$ following optical excitation at $t$=0 for the **a)** BTO/CTO 002 Bragg reflection, $l$=0, and the **b)** SL $l$=+1 and **c)** SL $l$=-1 reflections near the BTO/CTO 002 reflection. The intensity of each reflection is normalized to its peak value at $t < 0$.

The wavevectors of the maximum intensity for each reflection, $Q_{z,max}(l)$, vary as a function of $t$. The temporal variations of $Q_{z,max}(l)$ with $t$ for $l$=0,+1, and -1 reflections were extracted by fitting the peak profiles at each time step in the intensity maps in Figure 2 and are shown in **Figure 3**. The propagation of acoustic strain pulse to the surface and back to SL/SRO interface between 0 and 64 ps is responsible for the large variations in $Q_{z,max}(l)$ in this time regime. After 64 ps, $Q_{z,max}(l)$ maintains a lower value for the full-time range of the experiment, indicating that there is a photoinduced expansion that persists after the acoustic propagation is complete. The expansion in the 64-100 ps time regime can be unambiguously separated from thermal expansion because the BTO/CTO SL shows thermal contraction in the 25-100 °C temperature regime, as described in the Supporting Information.

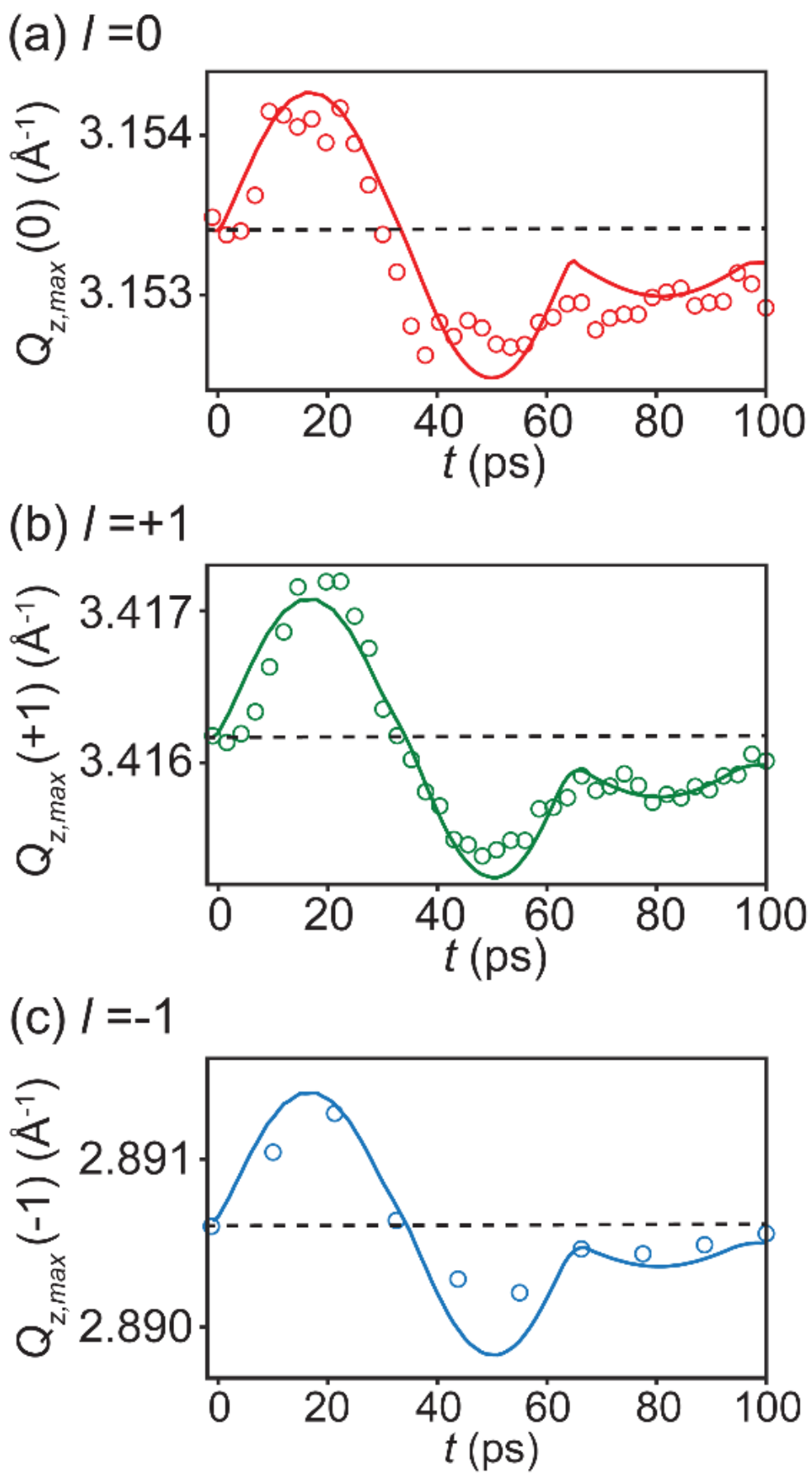

**Figure 3**. Measured (points) and simulated (solid line) time dependence of $Q_{z,max}(l)$ for **a)** BTO/CTO 002 Bragg reflection, $l$=0, and the **b)** SL $l$=+1 and **c)** SL $l$=-1 reflections near the BTO/CTO 002 reflection. The dashed lines indicate the steady-state values of $Q_{z,max}(l)$.

A comparison of the experimental results with an x-ray scattering simulation provides further insight into the structural distortion within the component layers. The atomic arrangement considered in the simulation consisted of an idealized BTO/CTO SL with time- and depth-dependent strain imposed by the acoustic pulse, depolarization-field-screening-driven expansion, and heating of SL. The diffracted intensity was calculated at each experimental time step using x-ray kinematical diffraction calculations. The initial atomic positions were set to provide equal ionic polarization in the BTO and CTO layers.

The strain imposed in the simulation consisted of the sum of the strain from the acoustic pulse, heating, and $\varepsilon_{net,depolarization}$, the depolarization-field-screening-driven average strain in the SL. The strain $\varepsilon_{net,depolarization}$ was further decomposed into the lattice distortion of the BTO and CTO components using $\varepsilon_{BTO,depolarization}$ and $\varepsilon_{CTO,depolariation}$, the out-of-plane component $\varepsilon_{33}$ of the strain tensor in the BTO and CTO layers of the SL, respectively. In terms of these components, the net strain is given by $\varepsilon_{net,depolarization}=\frac{1}{6}(2\ \varepsilon_{BTO,depolarization} + 4\ \varepsilon_{CTO,depolarization})$. The values of $\varepsilon_{BTO,depolarization}$ and $\varepsilon_{CTO,depolarization}$ were separately determined by comparing the experimental data with the intensity predicted by the diffraction simulation. The ratio of the depolarization-field-screening-driven distortions of the two components is $r = \frac{4\ \varepsilon_{CTO,depolarization}}{2\ \varepsilon_{BTO,depolarization}}$.

The values of $Q_{z,max}(l)$ were extracted from the simulated kinematical diffraction patterns for $l$=0, +1, and -1 at each time step and compared to the experimentally observed time dependence of $Q_{z,max}(l)$. Simulations with $\varepsilon_{net,depolarization}$=0.007 (±0.002)% agree with the experimental time dependence of $Q_{z,max}(l)$ for $l$=0,+1, and -1 and are plotted as solid lines in Figure 3a-c. The simulated variation of $Q_{z,max}(l)$ does not depend on the value of $r$. The simulations include the strain pulse launched due to optical absorption in the SRO layer with a magnitude of 0.7%. The depolarization-field-screening-driven strain $\varepsilon_{net,depolarization}$ is thus orders of magnitude less than the peak magnitude of the strain pulse from the SRO. The two contributions, the amplitude of the acoustic strain pulse and the depolarization-field-screening-driven strain, have their origins in different layers of the heterostructure (i.e. the SL and SRO) and there is thus no physical significance in their relative magnitudes.

The experimentally observed variation of the diffracted intensity was used to determine the value of $r$ and thus to find the component-specific contributions to the depolarization-field-screening-driven strain, $\varepsilon_{BTO,depolarization}$ and $\varepsilon_{CTO,depolariation}$. The time dependence of the integrated intensities of the $l$=0, +1, and -1 reflections are shown in **Figure 4**. The intensity of each reflection is normalized to its value at $t$<0. The intensities vary rapidly during the acoustic pulse propagation, between $t$=0 and 64 ps. The intensity does not, however, return to its initial value after the acoustic pulses have propagated into the substrate at $t$ > 64 ps. Simulated kinematical diffraction patterns with $\varepsilon_{net,depolarization}$=0.007 (±0.002)%, taken from the variation in $Q_{z,max}(l)$ in Figure 3, and $r$=-0.5 (±0.2) agree with the experimental data and are shown as solid lines in Figure 4a-c. The variation of the goodness of fit for different values of $\varepsilon_{net,depolarization}$ and $r$ and the determination of the uncertainty in these parameters are described in the Supporting Information.

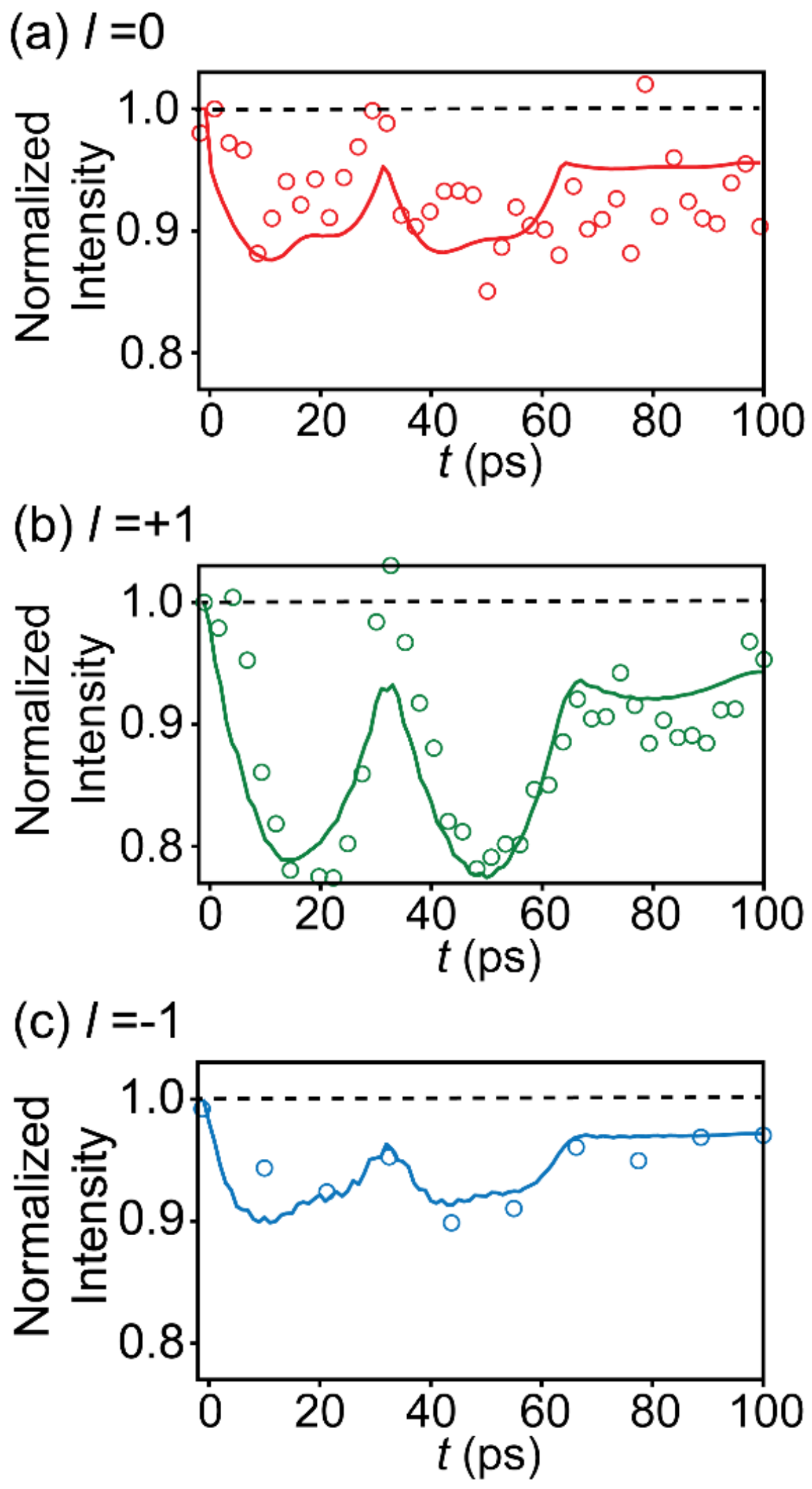

**Figure 4.** Time dependence of the measured (points) and simulated (solid lines) diffracted x-ray intensity for the **a)** BTO/CTO 002 Bragg reflection, *l*=0, and the **b)** SL *l*=+1 and **c)** SL *l*=-1 reflections near the BTO/CTO 002 reflection. The intensity of each reflection is normalized to its value at $t < 0$.

The negative value of $r$ obtained from the intensity data in Figure 4 indicates that depolarization-field screening produces opposite strains in BTO and CTO layers. In BTO thin films, screening of the depolarization field leads to an increase in the polarization of the BTO layers and hence the BTO layers have expansive strain.[9] A similar effect likely applies to superlattices, and we thus expect $\varepsilon_{BTO,depolarization}$ to be positive. The negative value of $r$ then indicates that there is a photoinduced compressive strain in the CTO layers. The fitted values of $\varepsilon_{net,depolarization}$= 0.007% and $r$ =-0.5 from Figures 3 and 4 give $\varepsilon_{BTO,depolarization}$ = 0.04% and $\varepsilon_{CTO,depolarization}$ = -0.01%.

In the LGD theory description, the relationship between the lattice parameter $c_i$ and polarization $P_i$ of layer $i$ is $P_i^2 \propto \frac{c_i}{a} - 1$.[13] With this approximation, the net change in polarization due to the optical excitation can be calculated from the values of $\varepsilon_{net,depolarization}$ and $r$. The in-plane lattice parameter $a$ is fixed by the epitaxial synthesis of the SL on the STO substrate, $a$=3.905 Å. The fractional change in the polarization of each component can be approximated as $\frac{\Delta P_i}{P_i} = \frac{1}{2}\varepsilon_i\left(\frac{c_{i,0}}{c_{i,0}-a}\right)$, where $c_{i,0}$ is the steady-state out-of-plane lattice parameter of component $i$ (see Supporting Information). The values of the steady-state out-of-plane lattice parameter of each component layer are assumed to be equal to the average lattice parameter of the SL measured from the 002 SL Bragg reflection at $l$=0, $c_{BTO,0} = c_{CTO,0}$ = 3.985 Å. This approach with the experimental results $\varepsilon_{net,depolarization}$= 0.007 (±0.002)% and $r$ = -0.5 (±0.2) give $\Delta P_{BTO}/P_{BTO}$= 2% and $\Delta P_{CTO}/P_{CTO}$ = -1%. The small fractional change in the polarization of BTO layers indicates that there is a partial screening of the depolarization field in BTO layers. Diffraction studies in an applied electric field indicated that the BTO and CTO components have equal piezoelectric strain.[26] The difference between the magnitudes of $\Delta P_{BTO}$ and $\Delta P_{CTO}$ is thus an intriguing result because screening of the depolarization field should normally be expected to create equal and opposite changes in the polarization and the strain in the two layers. The origin of the difference may arise from other contributions to the polarization of the CTO layers in SLs, including a reduction in the antipolar rotation of CTO oxygen octahedra at the BTO/CTO interface in comparison with bulk CTO.[17, 27-28]

The oxygen octahedral rotation is suppressed in the CTO layers near the BTO/CTO interfaces because BTO strongly resists oxygen octahedral rotation.[17, 27] The photoexcited charge carriers are not expected to affect the polarization contribution arising from the octahedral rotation suppression. Hence, the lattice contraction and corresponding polarization reduction in the CTO layers are smaller in magnitude compared to that of the BTO layers. The suppression of oxygen octahedral rotation, however, favors the polarization in [111] direction in the CTO layers, similar to $BiFeO_3$.[17] The complete compensation of the depolarization field in the BTO layers would nullify the contribution of the BTO polarization to the polarization of CTO layers along [001] direction. Therefore, if optical excitation with sufficiently high fluence to screen the depolarization field fully in the BTO were possible, a distinct metastable polarization configuration in the CTO layers could be produced. In the case of complete screening the CTO polarization would be dominated only by strain and octahedral rotation effects.

## Conclusion

The ultrafast response of BTO/CTO SLs to an above-bandgap optical pulse produces strain in each component layer. The screening of the depolarization field in the SL results in a reduction in the magnitude of the internal electric field in the CTO and leaves the CTO layers with reduced polarization. An analysis of photoinduced strain shows that the depolarization-field screening in the SL leads to a 2% increase in the polarization of BTO layers and a reduction of 1% in the polarization of CTO layers. The case of optical excitation is completely different than an applied electric field, as when the SL is incorporated in a thin-film capacitor. An applied electric field leads to piezoelectric expansion in both the BTO and CTO components, as expected from the equal steady-state polarization of these layers.[26] Optical excitation leads instead to a reduction in the polarization and lattice parameter of CTO, providing a different and complementary approach for the systematic variation of the polarization of the DE layers. The reduction of the polarization in the CTO layers along the [001] direction through depolarization-field screening opens up the possibility of attaining a metastable polarization in the CTO layers.

## Experimental Section

The BTO/CTO SL heterostructure consisted of 80 periods of the BTO/CTO repeating unit, with a SL thickness of 200 nm, epitaxially grown on an (001)-oriented 5-nm thick SRO layer on a (001) STO substrate. The structural distortion resulting from above-bandgap optical excitation was probed at the XSS beamline of Pohang Accelerator Laboratory XFEL (PAL-XFEL), using the arrangement in Figure 1b.[35] A 100 fs-duration π-polarized optical laser pump with 3.1 eV photon energy and optical fluence of 13.5 mJ cm$^{-2}$ was used to excite the SL. The x-ray pulses had a photon energy of 9.7 keV, 25 fs duration, and a repetition rate of 30 Hz. The x-ray fluence was selected to maintain the diffracted signal at an intensity within the dynamic range of the multi-port charge-coupled device x-ray detector and to be lower than the damage threshold previously measured for $BiFeO_3$ thin film layers under the same conditions.[36] The diffraction patterns corresponding to SL $l=0$, $l=+1$ and, $l=-1$ reflections near the 002 reflection were measured in the delay range -1 to 100 ps.

**Supporting Information**

Supporting Information is available.

**Acknowledgments**

The authors gratefully acknowledge support from the U.S. DOE, Basic Energy Sciences, Materials Sciences and Engineering Division, under contract no. DE-FG02-04ER46147. H.J.L. acknowledges support by the National Research Foundation of Korea under grant 2017R1A6A3A11030959. Y.J.A. acknowledges support from the National Science Foundation through grant number DMR-1609545. J.C. acknowledges support from the CNRS project GOtoXFEL. The authors acknowledge the use of characterization facilities supported by the National Science Foundation through the University of Wisconsin Materials Research Science and Engineering Center (DMR-1720415). The experiment was performed at the XSS beamline of PAL-XFEL (proposal no. 2018-2nd-XSS-016) funded by the Ministry of Science and ICT of Korea. The synthesis work at ORNL was supported by U.S. DOE, Basic Energy Sciences, Materials Sciences and Engineering Division.

**Conflict of Interest**

The authors declare no conflict of interest.

**Author Contributions**

D.S.G. and H.J.L. contributed equally to this work. H.J.L. and P.G.E. designed the experiments. H.J.L., Y.A., J.C., T.K., S.U., J.Y.L., S.H.C., S.K., I. E., M.K., S.Y.P., K.S.K., J.Y.J., and P.G.E. carried out the time-resolved free-electron-laser x-ray diffraction experiments. D.S.G. and H.J.L. analyzed the scattering data and developed the supporting calculations. H.N.L. synthesized the superlattice sample. D.S.G. and P.G.E wrote the manuscript. All authors contributed to the discussions and the revision of the manuscript.

Supporting Information

**Optically Induced Picosecond Lattice Compression in the Dielectric Component of a Strongly Coupled Ferroelectric/Dielectric Superlattice**

*Deepankar Sri Gyan, Hyeon Jun Lee, Youngjun Ahn, Jerome Carnis, Tae Yeon Kim, Sanjith Unithrattil, Jun Young Lee, Sae Hwan Chun, Sunam Kim, Intae Eom, Minseok Kim, Sang-Youn Park, Kyung Sook Kim, Ho Nyung Lee, Ji Young Jo, and Paul G. Evans*

## 1. Measurement of Longitudinal Acoustic Sound Velocity

The impulsive optical excitation at $t$=0 causes the scattered intensity near each Bragg reflection of the SL to oscillate at a frequency determined by the longitudinal acoustic phonon dispersion.[1] The relevant wavevector is $\Delta Q_z$, with $\Delta Q_z = Q_z - Q_{z,max}(l)$, where $Q_{z,max}(l)$ is the wavevector corresponding to the SL reflection with index $l$ at each time point and $Q_z$ is the wavevector at which the intensity is measured.

The measured time dependence of diffracted intensity corresponding to the thickness fringes near $l$=+1 reflection was compared to the simulation using values of the longitudinal sound velocity ranging from 5 to 7 km/s. Agreement between measured and simulated time dependance of intensities of the thickness fringes was observed for the sound velocities in the range of 5.7-6.1 km/s. We concluded that the $v_{SL}$ in the BTO/CTO SL is 5.9 (±0.2) km/s.

A separate measurement of $v_{SL}$ was obtained using the frequency-wavevector dispersion derived from the time-domain Fourier transform of the scattered intensity as a function of $\Delta Q_z$. **Figure S1** shows the amplitude of the time-domain Fourier-transform of the intensity for the SL $l$=+1 reflection near the BTO/CTO SL 002 reflection using the measured intensity in Figure 2b.

The longitudinal frequency dispersion obtained from the Fourier transform of the intensity map shows a linear dispersion up to $\Delta Q_z$=±0.015. A linear wavevector-frequency dispersion corresponding to $v_{SL}$=5.9 km/s is shown in Figure S1.

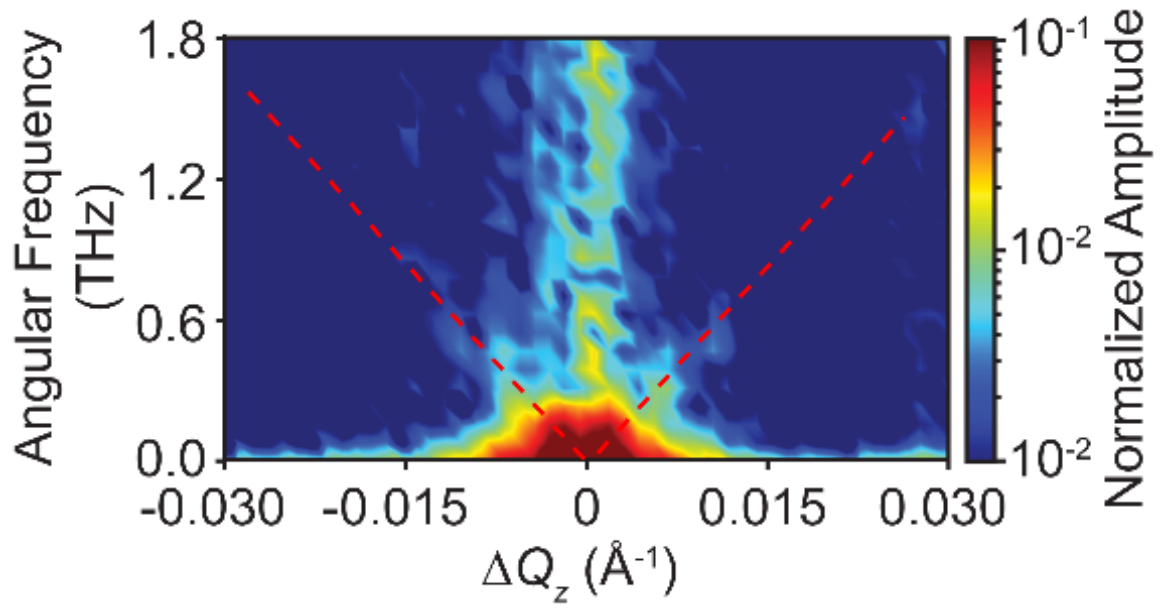

**Figure S1.** Time-domain Fourier transform of the diffracted x-ray intensity near the SL $l$=+1 reflection near the 002 Bragg reflection. The amplitude of the transform is normalized to 1 at its maximum at zero angular frequency and $\Delta Q_z$=0. The dashed line represents the linear wavevector-frequency dispersion with $v_{SL}$ = 5.9 km s$^{-1}$.

The longitudinal acoustic sound velocity in the SL can be predicted by treating the SL as an elastic continuum. In this approach, the predicted longitudinal acoustic sound velocity $v_{SL,predicted}$ is:[2]

$$v_{SL,predicted} = (d_{BTO}+d_{CTO})\left[\frac{d_{BTO}^2}{v_{BTO}^2}+\frac{d_{CTO}^2}{v_{CTO}^2}+\left(x+\frac{1}{x}\right)\frac{d_{BTO}d_{CTO}}{v_{BTO}v_{CTO}}\right]^{-\frac{1}{2}} \quad \text{(S1)}$$

Here $d_i$ is the thickness of layer $i$, $v_i$ is the longitudinal acoustic sound velocity in layer $i$, $x=(\rho_{CTO}/v_{CTO})/(\rho_{BTO}/v_{BTO})$, and $\rho_i$ is the mass density of layer $i$. The values of the parameters appearing in equation S1 are given in **Table S1**. The sound velocity in the SL predicted using equation S1 is 7.8 km s$^{-1}$.

The sound velocity calculated using the elastic continuum approximation (7.8 km s$^{-1}$) is higher than the measured value of 5.9 km $s^{-1}$, possibly because the continuum does not consider the effect of interfaces between the component layers.[3] For 80 repeating units of BTO/CTO SL used in the experiment, the BTO/CTO interfaces will have a significant effect on acoustic propagation and are thus expected to be smaller than the calculated value.

| | |
|---|---|
| BTO repeating unit thickness ($d_{BTO}$) | 8.17 Å (2 u. c.) |
| CTO repeating unit thickness ($d_{CTO}$) | 15.55 Å (4 u. c.) |
| BTO longitudinal sound velocity ($v_{BTO}$) | 6.5 km $s^{-1}$ [4] |
| CTO longitudinal sound velocity ($v_{CTO}$) | 8.8 km $s^{-1}$ [5] |
| BTO mass density ($\rho_{BTO}$) | 5.95 g $cm^{-3}$ [6] |
| CTO mass density ($\rho_{CTO}$) | 3.94 g $cm^{-3}$ [6] |

**Table S1:** Parameters for the calculation of sound velocity using an elastic continuum model.

## 2. Temperature Dependence of SL Lattice Parameter

The thermal expansion of the BTO/CTO SL was studied using x-ray diffraction with a laboratory x-ray source. **Figure S2a** shows the fractional change of the out-of-plane lattice parameter as a function of temperature measured using the $l$=0, $l$=+1, and $l$=-1 reflections of the SL and the 002 reflection of the STO substrate. The coefficient of thermal expansion (CTE) for STO measured using a linear fit to the experimental data in Figure S2a is $9.9 \pm 0.3 \times 10^{-6}$ $K^{-1}$. The measured CTE of STO matches the reported value in this temperature range, $9.6 \times 10^{-6}$ $K^{-1}$.[7]

The variation of the lattice parameter of the SL as a function of temperature is affected by the epitaxial constraint imposed by the substrate, which remained at room temperature during the optically excited experiments but was heated during the laboratory experiments. The ultrafast XFEL diffraction experiments were conducted under conditions in which the SL and SRO layers were free to expand or contract along the surface-normal direction but were clamped to the in-plane lattice parameter of the room-temperature STO substrate. The effective coefficient of thermal expansion under the conditions of the optically pumped experiment was obtained by accounting for the stress imposed by the substrate to maintain constant in-plane lattice parameters in the SL and SRO layers. The additional out-of-plane strain arising from distorting the in-plane lattice parameter of the heated SL and SRO layers to match the room-temperature substrates is:

$$\varepsilon_{out} = \frac{2\nu}{1-\nu}\varepsilon_{in}.$$

Here $\varepsilon_{out}$ is the out-of-plane strain imposed by the in-plane strain $\varepsilon_{in}$ and ν is the Poisson ratio. The Poisson ratios of BTO and CTO are 0.23 and 0.26, respectively.[6] For the BTO/CTO SL, the value of ν of SL was taken to be the volume average of these values, 0.25.[8]

**Figure S2b** shows the fractional change of the out-of-plane lattice parameter of the SL relative to room temperature under elastic conditions corresponding to a room-temperature STO substrate and a heated film. The CTE of the SL in the temperature range 25 to 100 °C under these this constraint, measured by averaging the CTE values measured from SL *l*=0, *l*=+1, and *l*=-1 reflections, is $-5 \times 10^{-7}$ $K^{-1}$. The constraint imposed by the substrate causes the CTE of the BTO/CTO layer to have a smaller magnitude because the in-plane compression caused by the constant lattice parameter of the STO substrate results in an out-of-plane expansion, partially canceling the out-of-plane contraction that would occur with a heated substrate.

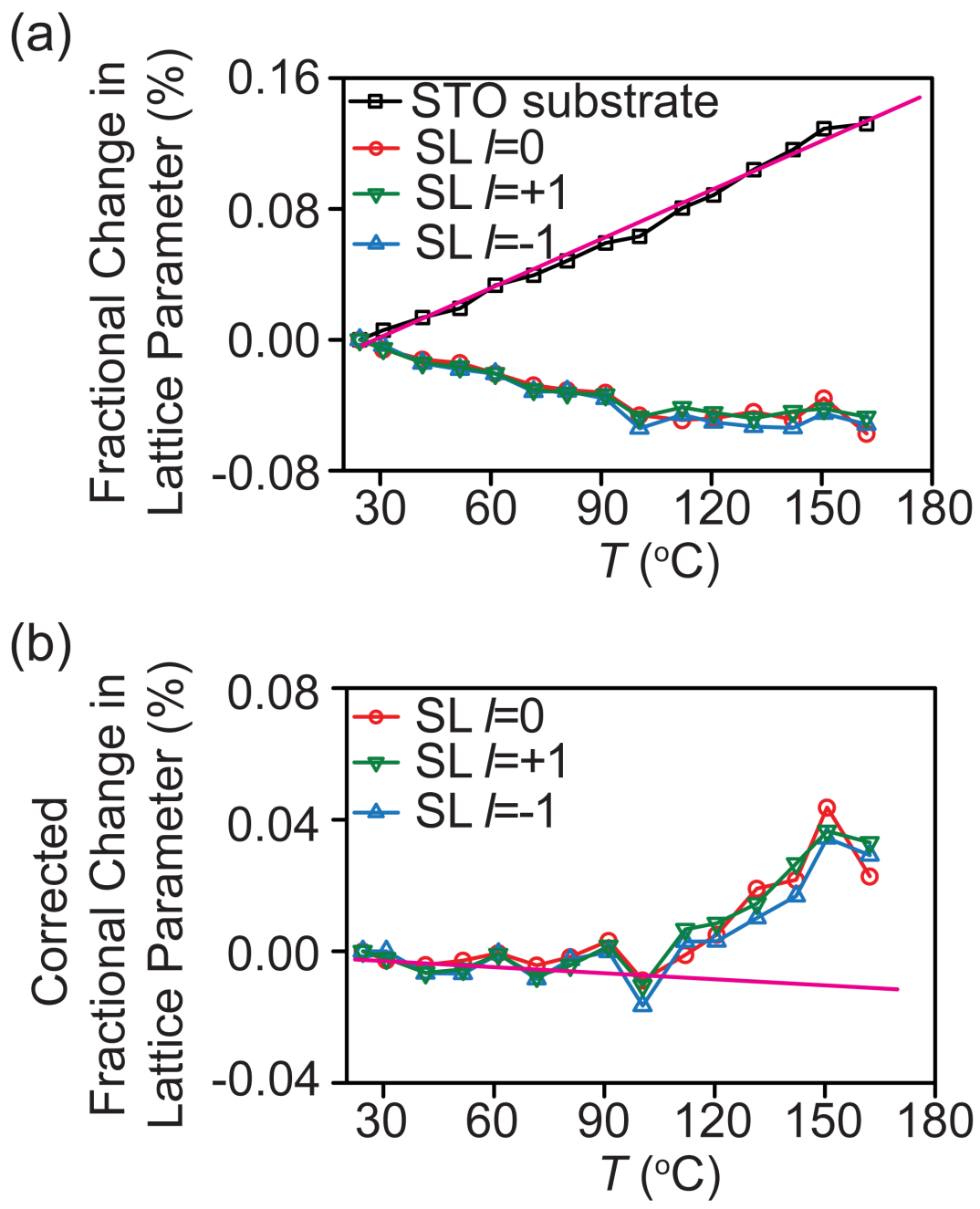

**Figure S2. a)** Temperature dependence of the fractional change in lattice parameter of the STO substrate and BTO/CTO SL, measured with the STO 002 reflection and the SL *l*=0, +1, and -1 reflections. The solid line shows a linear fit to the thermal expansion of the STO substrate. **b)** Fractional change in lattice parameter of SL, measured using the SL *l*=0, +1, and -1 reflections including the elastic correction to account for the room-temperature STO substrate. The solid line corresponds to a coefficient of thermal expansion of $-5 \times 10^{-7}$ $K^{-1}$ in the temperature range of 25-100 °C.

### 3. Temperature Effects in X-ray Diffraction Simulation

The effect of the increase in the temperature of the SL was incorporated in the diffraction simulation by including (i) the thermal contraction of the SL in the relevant temperature range, as in Figure S2b, and (ii) a scaling factor accounting for the temperature dependence of diffracted intensity.

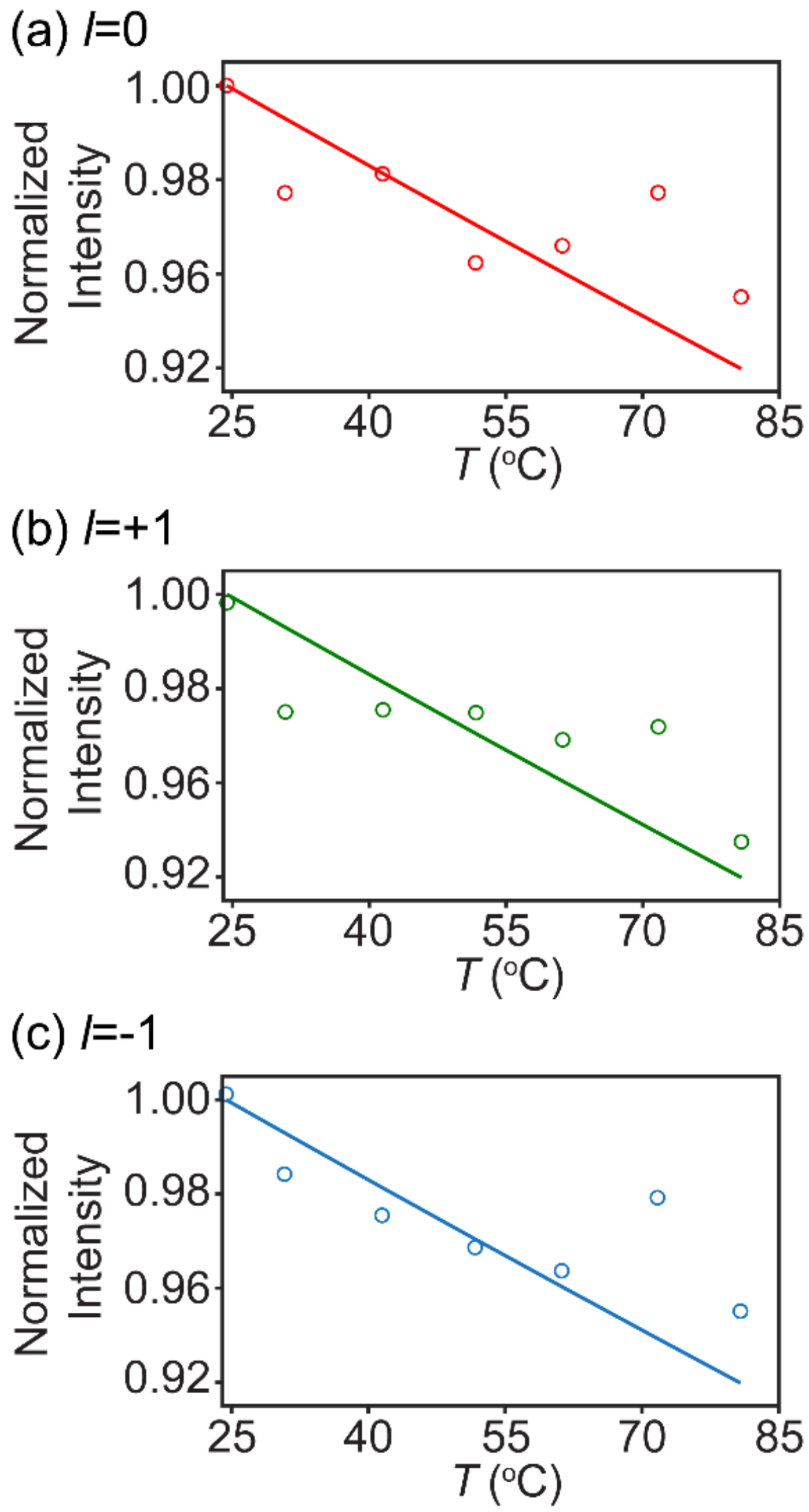

**Figure S3.** Temperature dependence of diffracted intensity (points) and trend lines for $2m=1.0 \times 10^{-3}$ (line) for **a)** $l$=0, **b)** $l$=+1, and **c)** $l$=-1 SL reflections.

The factor accounting for the temperature dependence of diffracted intensity was determined from the laboratory-source diffraction measurements. The intensity of the SL $l$=0, +1, and -1 reflections as a function of temperature is shown in **Figure S3**. The diffracted intensities were normalized to their values at room temperature value and then fitted with $I_{normalized}=\exp(-2m(T-T_0))$, where $T_0$ is room temperature. The coefficient $m$ arises from the Debye-Waller effect and accounts for the dependence of diffracted intensity on temperature. The solid lines in Figure S3 shows trendlines for the exponential model for SL $l$=0, +1, and -1 reflections with $2m = 1.0 \times 10^{-3}$. The effect of the temperature increase was incorporated in the simulations by multiplying the structure factor of each unit cell by $\exp(-m(T-T_0))$.[9] Note that this approximation is applicable when the changes in temperature are small for temperatures near the Debye temperature, which is 400-500 K for $BaTiO_3$ and 760 K for $CaTiO_3$.[10-11]

**4. Optical Absorption in the SL and SRO**

The incident optical pump fluence at the surface of the SL is $I_{0,SL}=I_0$, where $I_0$ is the incident fluence. The reflectivity of the BTO/CTO SL measured during the experiment with incident pulses with a fluence of 13.5 mJ cm$^{-2}$ was $R_{SL} = 0.16$.

The optical properties of the BTO/CTO SL have not yet been reported in detail. We adopt an approximation based on the normal incidence optical constants, which preserves the key features of the problem, including the far larger optical absorption per unit depth in the SRO bottom electrode. Within this approach, the optical absorption length $\zeta_{SL}$ for the SL was calculated using an effective medium approximation:[8]

$$\zeta_{SL} = \frac{\lambda}{4\pi k_{eff}}$$

with

$$k_{eff} = \sqrt{\frac{k_{BTO}^2 d_{BTO} + k_{CTO}^2 d_{CTO}}{d}}$$

Here, $\lambda$ = 400 nm is the vacuum wavelength of the incident optical pulse, $k_{eff}$ is the effective extinction coefficient of the SL, and $k_{BTO}$ =0.023[12] and $k_{CTO}$=0.025[13] are the optical extinction coefficients of BTO and CTO. The energy absorbed per unit volume in the SL is $w_{abs,\mathrm{SL}}$, given by:[14]

$$w_{abs,SL}(z) = \frac{(1 - R_{SL})I_{0,SL}}{\zeta_{SL}} e^{-\frac{z}{\zeta_{SL}}}$$

Here $z$ is the depth from the surface of the SL.

The optical absorption length in the SRO bottom electrode was $\zeta_{SRO}$ = 30 nm.[15] The incident fluence for the SRO electrode was calculated using $I_{0,SRO} = (1 - R_{SL})I_0 e^{-\frac{d}{\zeta_{SL}}}$, where $d$ is the thickness of the SL. The energy absorbed per unit volume inside the SRO electrode is $w_{abs,SRO}$ given by:[14]

$$w_{abs,SRO}(z_{SRO}) = \frac{(1 - R_{SRO})I_{0,SRO}}{\zeta_{SRO}} e^{-\frac{z_{SRO}}{\zeta_{SRO}}}$$

Here, $z_{SRO}$ is the depth within the SRO electrode relative to the SL/SRO interface and $R_{SRO}$ is the optical reflectivity of the SRO/SL interface. We used $R_{SRO}$ =0.2, the reflectivity of SRO in vacuum.[15] The optical absorption in the STO is far lower than the SRO, with an extinction coefficient of 0.0635 at 400 nm.[12] In addition, STO lacks the depolarization screening induced distortion mechanism that enables a depolarization-field-screening driven structural response in the SL. The effects of optical absorption in the STO substrate have thus not been considered in these calculations.

The calculated optical absorption profile is shown in **Figure S4**. The fractions of the optical fluence absorbed in the SL and the SRO bottom electrode are 11% and 10%, respectively.

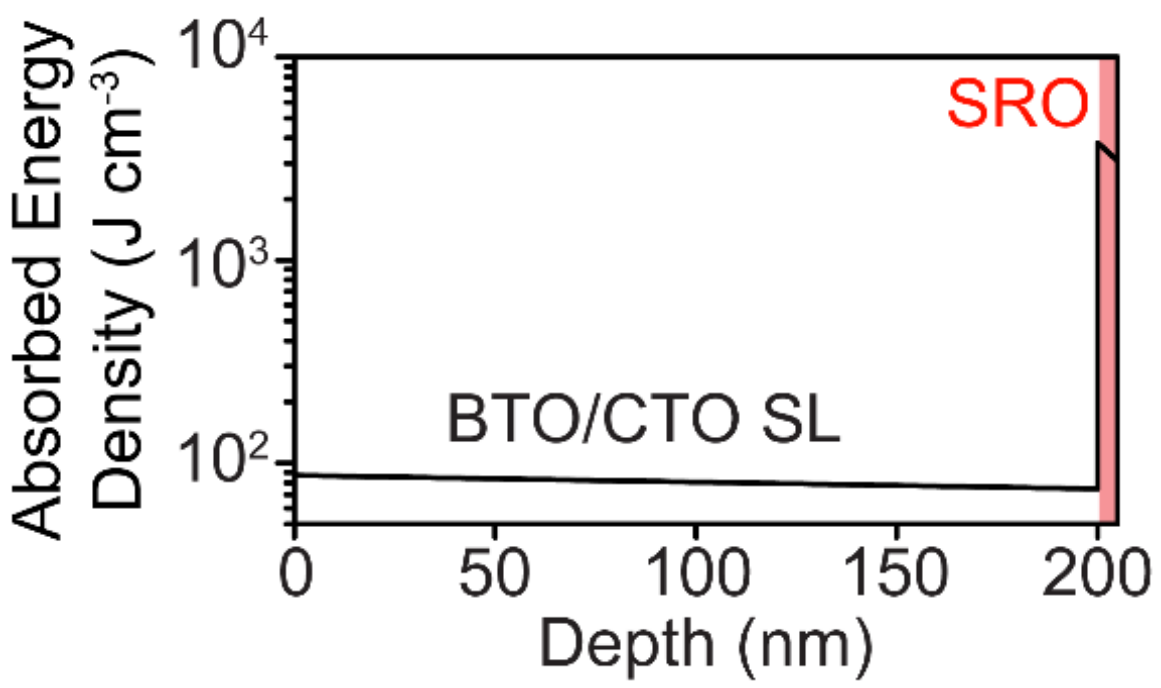

**Figure S4.** Depth profile of energy absorbed per unit volume by BTO/CTO SL and SRO bottom electrode for 13.5 mJ cm$^{-2}$ incident fluence.

## 5. Temperature Increase in the SL due to Heat Transfer from the SRO Layer

The increases in the temperature of the SL and SRO were determined by solving the time-dependent thermal diffusion equation in the time interval following the optical excitation. The simulations employed an optical fluence of 13.5 mJ cm$^{-2}$, matching the experiment. The simulation is thus expected to be valid in the time before the energy absorbed in the SRO is distributed through the thickness of the SL. The initial increase in the temperature of the SRO was:[14]

$$T_{SRO}(z_{SRO}) = \frac{w_{abs,SRO}(z_{SRO})}{c_{p,SRO}\ \rho_{SRO}}$$

Here, $w_{abs,SRO}$ is the energy per unit volume deposited in the SRO electrode, $c_{p,SRO}$ = 180 J mol$^{-1}$ K$^{-1}$ [16] and $\rho_{SRO}$ =6.20 g cm$^{-3}$ [15] are the specific heat and the mass density of SRO electrode. This temperature rise corresponds to the thermal state of SRO at $t$=0, just after the optical excitation.

The temperature as a function of depth and time $T(z,t)$ in the SL was determined using the heat diffusion equation:

$$\rho c_p \frac{\partial T(z,t)}{\partial t} = \kappa \frac{\partial^2 T(z,t)}{\partial z^2}$$

Here, $\kappa$ is the thermal conductivity. The values of the parameters are different for SL, SRO, and STO. There are two interfaces, i.e., the SL/SRO interface and the SRO/STO interface and four interface thermal boundary conditions:

At the SL/SRO interface:

$$-k_{SL} \frac{\partial T_{SL}}{\partial z} = h_{SL/SRO}(T_{SL} - T_{SRO}) \qquad \text{for } z=d_{SRO}^-$$

$$-k_{SRO} \frac{\partial T_{SRO}}{\partial z} = h_{SL/SRO}(T_{SL} - T_{SRO}) \qquad \text{for } z=d_{SRO}^+$$

At the SRO/STO interface:

$$-k_{SRO} \frac{\partial T_{SRO}}{\partial z} = h_{SRO/STO}(T_{SRO} - T_{STO}) \qquad \text{for } z=d_{STO}^-$$

$$-k_{STO} \frac{\partial T_{STO}}{\partial z} = h_{SRO/STO}(T_{SRO} - T_{STO}) \qquad \text{for } z=d_{STO}^+$$

Here, $h_{SL/STO}$ *and* $h_{SRO/STO}$ are the interfacial thermal conductances of the SL/SRO and SRO/STO interfaces, respectively and $d_{SRO}$ and $d_{STO}$ are the depths of SL/SRO interface and SRO/STO interfaces. The SRO/STO interfacial conductance is 0.8 GW $m^{-2}$ $K^{-1}$.[17] The value of $h_{SL/SRO}$ has not yet been reported, so we have used the value of $h_{SRO/STO}$ for the $h_{SL/SRO}$ interface.

The heat transfer equation was solved using the 1D forward Euler's method with a delay time step of 5 fs. A set of 1000 one-dimensional cells spanned the SL/SRO/STO heterostructure with a cell size of 1 nm. The temperature distribution immediately after optical excitation was taken as the initial conditions for the heat transfer simulation, as shown in **Figure S5a**. The initial temperatures of the SL film and SRO electrode were 50 °C and 750 °C respectively. The temperature distribution as a function of time and depth is shown in **Figure S5b.**

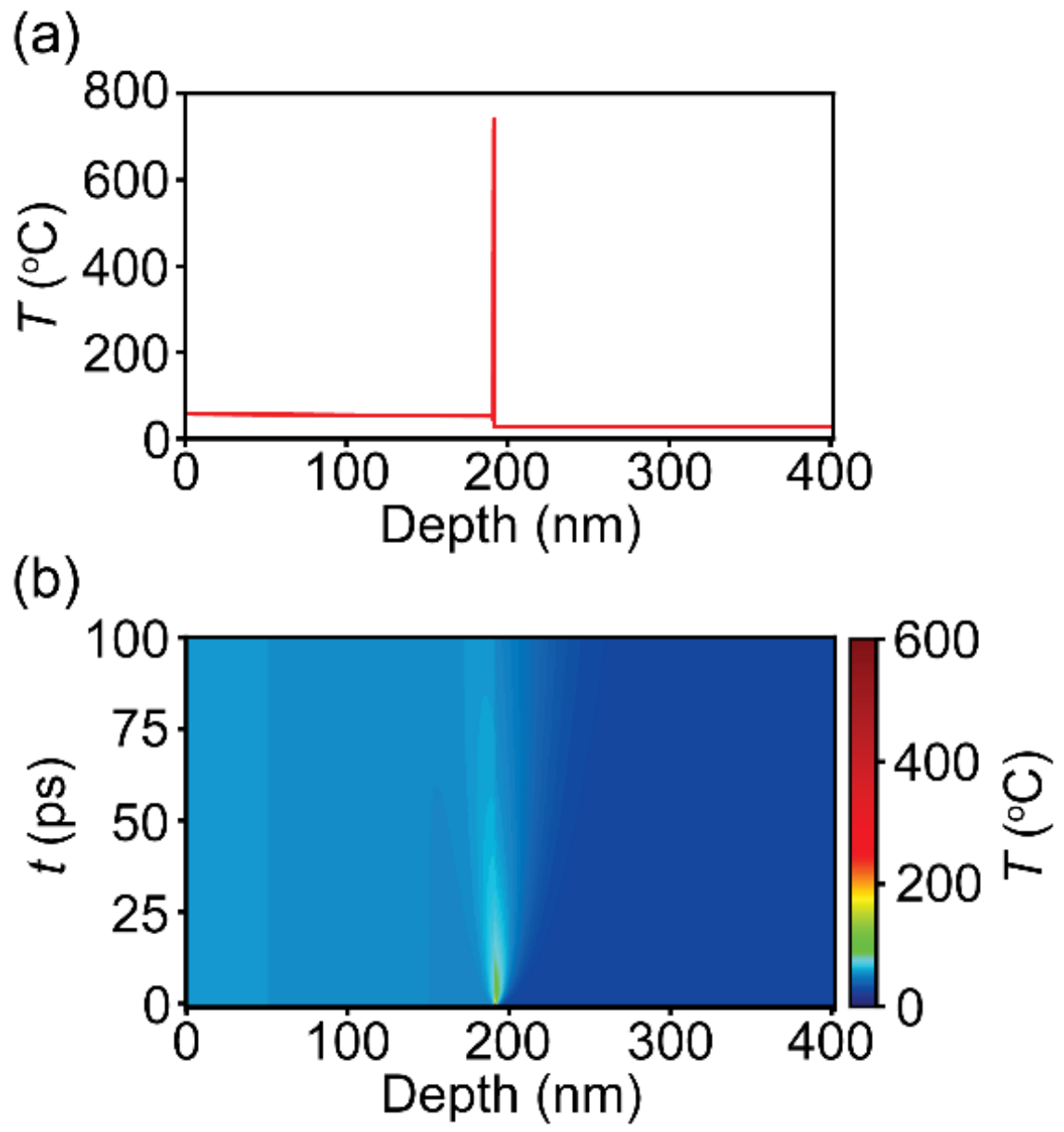

**Figure S5. a)** Temperature profile in the SL and the SRO electrode immediately after optical absorption. **b)** Variation of temperature with time and depth for 13.5 mJ cm$^{-2}$ fluence. The surface of the SL is at zero depth.

### 6. Simulation of Acoustic Pulses

We first consider the acoustic pulse with a small amplitude that arises from optical absorption in the SL. The out-of-plane strain $\varepsilon_{33}$ due to the acoustic pulse resulting from the impulsive heating of the SL layer is given by:[14]

$$\varepsilon_{33} = (1 - R_{SL})\frac{I_0\beta}{\zeta_{SL}C}\frac{1+\nu}{1-\nu}[e^{-\frac{z}{\zeta_{SL}}}\left(1 - \frac{1}{2}e^{-\frac{v_{SL}t}{\zeta_{SL}}}\right) - \frac{1}{2}e^{-\frac{|z-v_{SL}t|}{\zeta_{SL}}}sgn(z - v_{SL}t)]$$

Here $C=\rho c_p$ and $\beta$ is the coefficient of thermal expansion.

The calculated strain profile of the acoustic pulse arising from optical absorption in the SL at is shown for an arbitrarily selected snapshot at $t$=7 ps in **Figure S6a**. Figure S6a also shows the strain profile in the SL at $t$=7 ps due to the heating of the SL corresponding to the thermal transport from the heated SRO layer. The strain profile at $t$=7 ps of the acoustic pulse generated in the SRO electrode is shown in **Figure S6b**. Profiles at other times (not shown) are consistent with the propagation of the acoustic pulse but are otherwise similar to those in Figure S6. The strain due to optically induced heating of the SL and the heat transfer from the SRO layer to the SL, also shown in Figure S6a, are very small, of the order of 0.001%, compared to the strain pulse generated from the SRO layer.

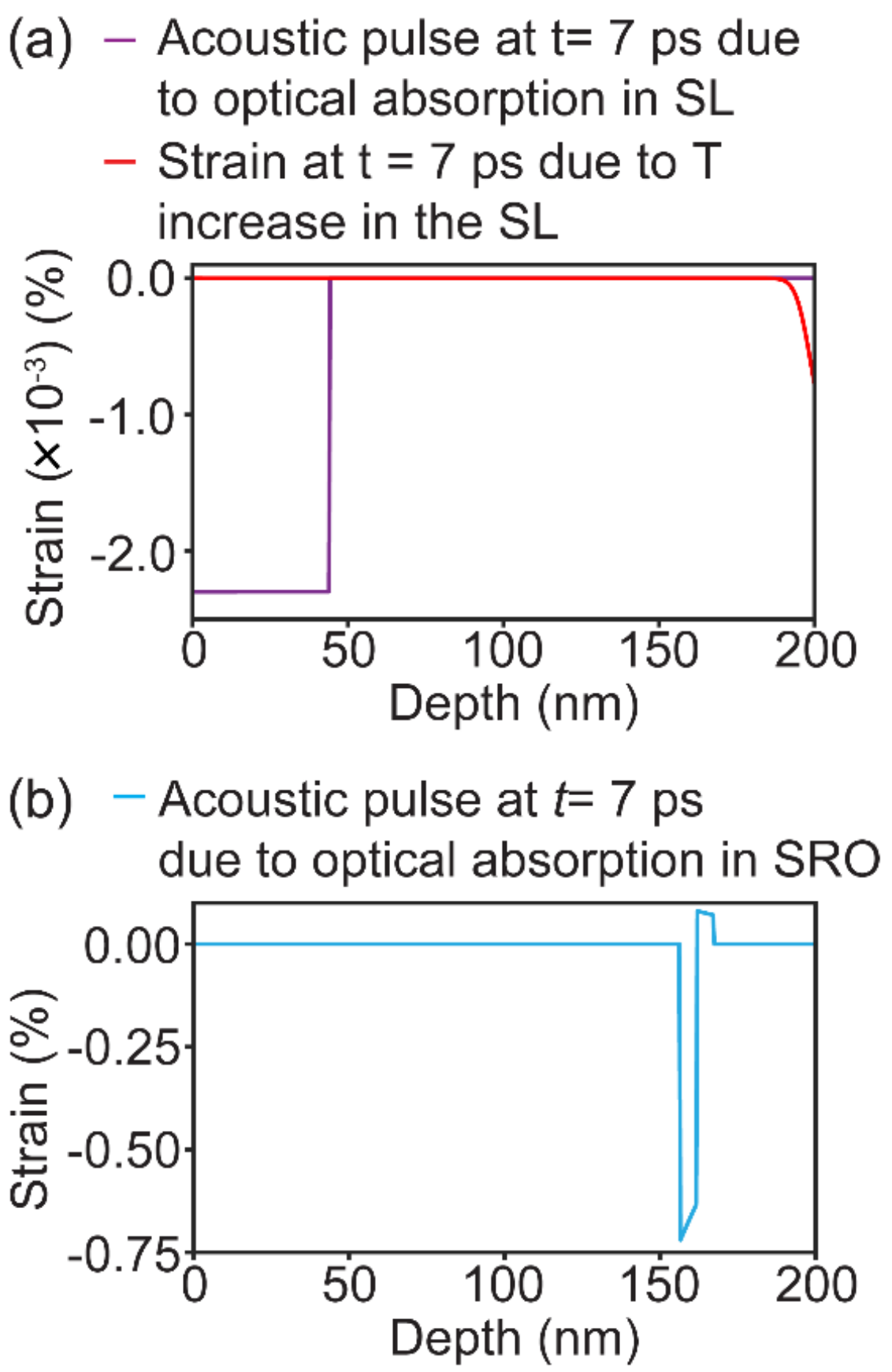

**Figure S6. a)** Snapshot of the strain profile of the acoustic pulse inside the SL due to optical absorption in SL and strain arising from the temperature increase of SL corresponding to the thermal transport from the heated SRO layer. The profiles are plotted for an arbitrarily selected delay time *t*=7 ps. **b)** Strain at *t*=7 ps in acoustic pulse launched by SRO bottom electrode.

The acoustic pulse launched from bottom SRO electrode has a far larger amplitude, on the order of 1%. The acoustic strain pulses propagate through the SRO layer and into the substrate after 64 ps and a smaller fraction of the acoustic pulses is reflected in the SL film. The amplitude of the acoustic pulse reflected at SL/SRO interface was calculated using the acoustic impedance mismatch between the SL and the SRO layer. The impedance ratio is:[18]

$$\frac{Z_{SL}}{Z_{SRO}} = \frac{\rho_{SL} v_{SL}}{\rho_{SRO} v_{SRO}}$$

Here, $Z_i$ $\rho_i$ and $v_i$ are the acoustic impedance, density, and sound velocity in layer *i*. Values of these parameters are given in **Table S2.** The ratio of the amplitude of the reflected pulse to the amplitude of the acoustic pulse propagating through the SL is $(Z_{SRO}-Z_{SL})/(Z_{SRO}+Z_{SL})$. The value of this ratio is 20% for the parameters in Table S2.

| | |
|---|---|
| SL mass density ($\rho_{SL}$) | 4.54 g cm$^{-3}$ (calculated in this work) |
| SL sound velocity ($v_{SL}$) | 5.9 km s$^{-1}$ (measured in this work) |
| SRO mass density ($\rho_{SRO}$) | 6.20 g cm$^{-3}$ [19] |
| SRO sound velocity ($v_{SRO}$) | 6.3 km s$^{-1}$ [19] |

**Table S2**: Parameters for the calculation of acoustic impedance mismatch.

### 7. Determination of Values of $\varepsilon_{net}$ and *r* by Evaluating the Root-Mean-Square Error

We have introduced two fitting parameters: (i) the total depolarization field screening driven strain $\varepsilon_{net,depolarization}=\frac{1}{6}(2\ \varepsilon_{BTO,depolarization} + 4\ \varepsilon_{CTO,depolarization})$ and (ii) the ratio of the depolarization field screening driven strains the two components $r = \frac{4\,\varepsilon_{CTO,depolarization}}{2\,\varepsilon_{BTO,depolarization}}$. The goodness of the fit to the data in Figure 3 and 4 was determined using the values of the root-mean-squared error (RMSE). The RMSE was computed in the time regime $t$ > 64 ps, where the depolarization-field screening effects are apparent. The RMSE is the root-mean-square difference between the points in the experimental measurement and the corresponding points of the simulated diffraction pattern.[20] A lower RMSE indicates that there is a better agreement between the experiment and the simulation.

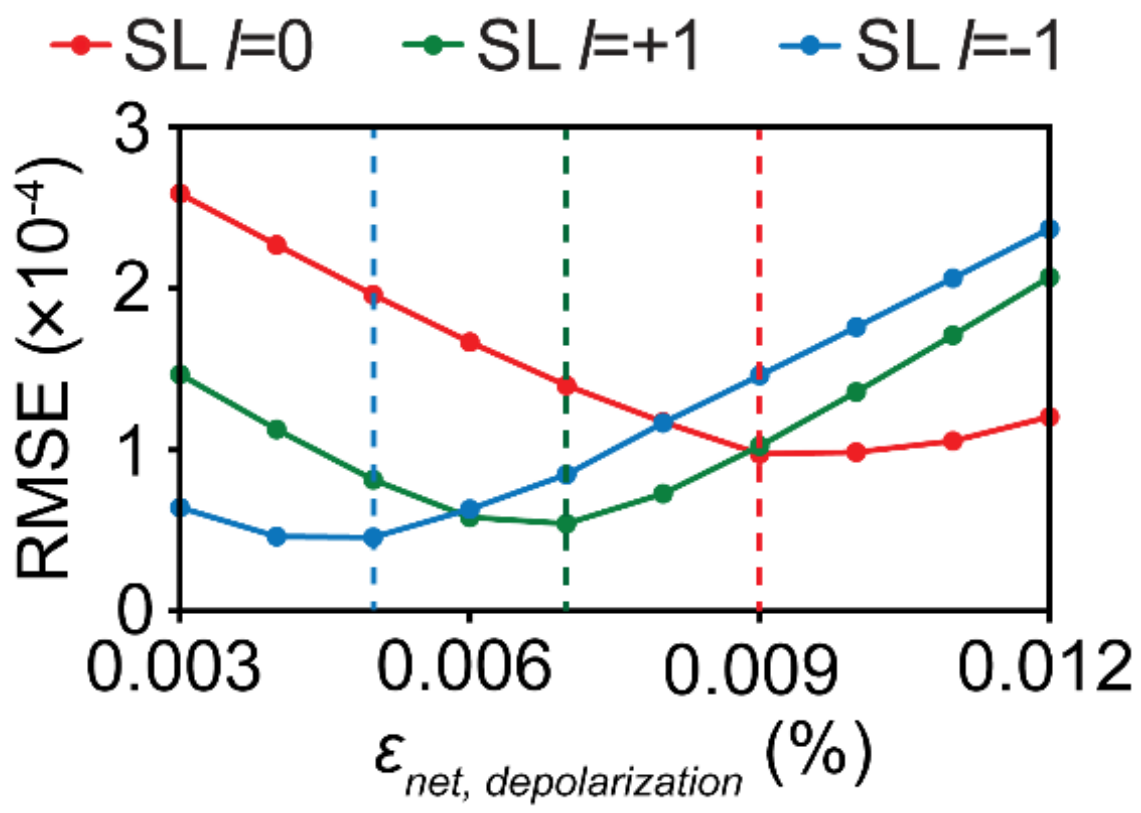

**Figure S7**. Dependence of the RMSE for $\varepsilon_{net,depolarization}$ using the wavevector data in Figure 3 for the SL $l$=0, SL $l$=+1, and $l$=-1 reflections. The fit was evaluated at the values of $\varepsilon_{net,depolarization}$ plotted as data points.

**Figure S7** shows the variation of the RMSE as a function of the strain $\varepsilon_{net,depolarization}$, obtained by a comparison of the model with the wavevector data in Figure 3. For the SL $l$=0, $l$=+1 and $l$=-1 reflections, the lowest RMSE was obtained for $\varepsilon_{net,depolarization}$=0.009%, 0.007%, and 0.005%, respectively. The value of the net photoinduced strain in the superlattice is thus in the range of 0.005% to 0.009% strain. The value of $\varepsilon_{net,depolarization}$ reported in the text is 0.007

(±0.002)%.

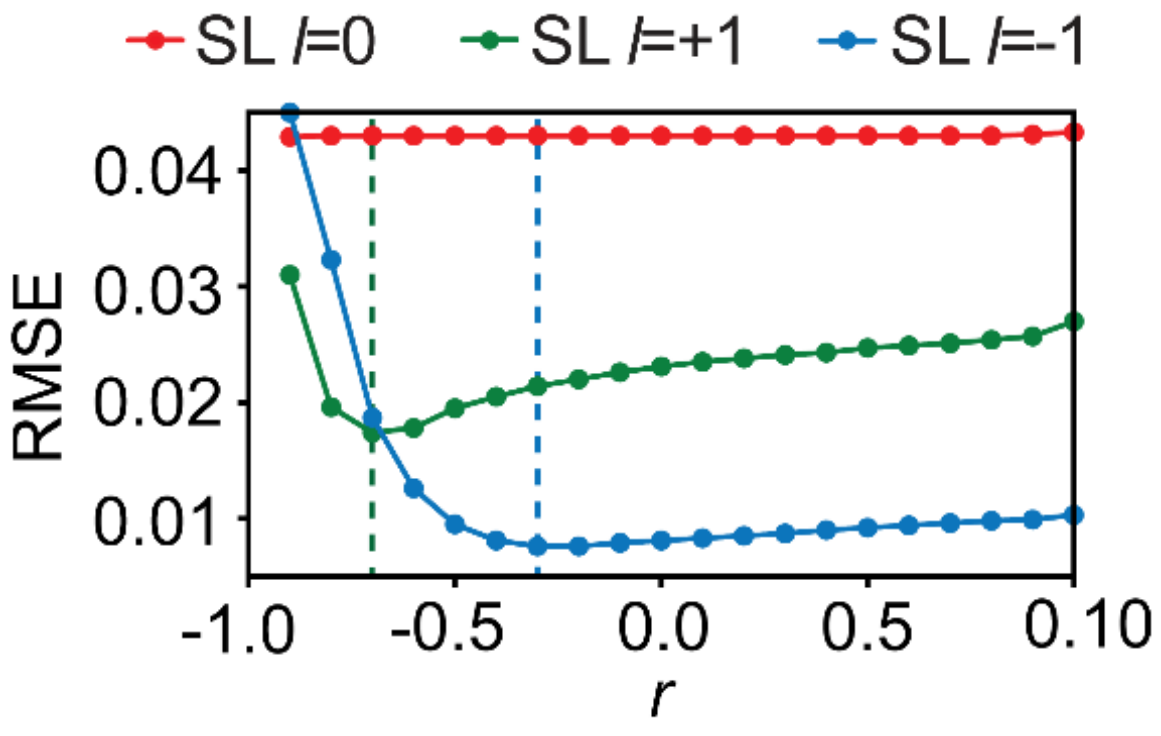

**Figure S8**. RMSE from the fitting of the intensity data in Figure 4 for the SL *l*=0, SL *l*=+1, and *l*=-1 reflections.

The value of the strain ratio *r* was determined by setting $\varepsilon_{net,depolarization}$=0.007% and considering different values of *r*. The values of the RMSE corresponding to different values of *r* are shown in **Figure S8**. The intensity of SL *l*=0 reflection is independent of the component-specific strain. Hence, for the SL *l*=0 reflection, the RSME remains constant throughout the range of values of *r*. For the SL *l*=+1 reflection, the best values of RMSE were observed for *r*=-0.7. For SL *l*=-1 reflection, the RMSE initially decreases with the value of *r* and starts to increase slowly after -0.3. There are only four data points for the *l*=-1 reflection the range *t*>64 ps. We thus observe a smaller change in the value of RSME as the value of *r* is increased. Comparing the values of RMSE for SL *l*=+1 and -1 reflections, we can conclude that the range of *r* for the best fit to the experimental data will lie in the range of -0.7 to -0.3. The value of *r* reported in the text is 0.5 (±0.2).

8. **Derivation of ΔP/P**

The relationship between strain and polarization in ferroelectrics is often described using the Landau-Ginzburg theory, which gives $P_{z,i}^2 \propto \left(\frac{c_i}{a} - 1\right)$.[21] Here $P_{z,i}$ is the ferroelectric polarization of component $i$ along the out-of-plane direction, $c_i$ is the out-of-plane lattice parameter of layer $i$, and $a$ is the in-plane lattice parameter. The fractional change in the polarization of component $i$ along the $z$-direction was calculated by differentiating the logarithm of the expression of polarization.

$$\mathrm{d}\left(2\ln\left(P_{z,i}\right)\right) = \mathrm{d}\left(\ln\left(\frac{c_i - a}{a}\right)\right)$$

Note that $\Delta a$=0 and the constant of proportionality does not change.

$$2\frac{\mathrm{d}P_{z,i}}{P_{z,i}} = \frac{a}{c_i - a}\frac{\Delta c_i}{a}$$

$$2\frac{\mathrm{d}P_{z,i}}{P_{z,i}} = \frac{c_i}{c_i - a}\frac{\Delta c_i}{c_i}$$

So,

$$\frac{\Delta P_{z,i}}{P_{z,i}} = \frac{1}{2}\varepsilon_i\left(\frac{c_i}{c_i - a}\right).$$